\documentstyle[twocolumn,prb,floats,psfig,aps]{revtex}

\begin{document}

\draft

\voffset 0.8cm

\oddsidemargin -1cm

\twocolumn[\hsize\textwidth\columnwidth\hsize\csname 
@twocolumnfalse\endcsname

\title{THERMODYNAMICS OF VORTEX LINES IN LAYERED \\
SUPERCONDUCTORS}

\author{Roberto Iengo and Claudio A. Scrucca \\ 
{\em International School for Advanced Studies,
Via Beirut 4, 34014 Trieste (Italy)} \\
{\em and INFN - Sezione di Trieste (Italy)}}

\date{\today}

\maketitle

\begin{abstract}

We study the dissipative thermodynamics of vortex lines in layered
superconductors within a simple string model in the dilute limit of
negligible vortex interactions and compute the specific heat $C_v$ in 
presence of arbitrary dissipation. The interplay of dissipation, inertia
and elasticity is shown to control the qualitative thermodynamical behavior
and their relative amount determines two very distinct regimes for the
specific heat. In the dissipation dominated case we find a behavior $C_v 
\sim \sqrt{T}$ for a large interval of temperature below $T_c$.

\end{abstract}


]

\section{Introduction}

As well known, type-II superconductors in a magnetic field exhibit an
intermediate mixed phase characterized by vortex condensation and partial
magnetic flux penetration through them. The dynamics of the latter crucially
drives the electromagnetic behavior, and has been extensively studied in the
last years \cite{Blat1}. It has been shown that under appropriate
assumptions, in particular for small magnetic field and vortex density, 
the elementary vortex  excitations in thin layered samples with transverse 
magnetic field are flux lines crossing each layer only once, joining pancake 
vortices forming in each layer and moving in the xy plane. 
A well established and studied continuum model \cite{Blat1}
exist for these string-like excitations and it has 
been succesfully applied to analyze both the dynamical 
\cite{Blat2,Ivlev,Blat3} and therodynamical \cite{Bula1,Bula2,Blat4} 
behavior.

In the following we study the specific heat of vortex lines within a simple
model which can be treated exactly and captures all the essential features
of vortex dynamics relevant in layered superconductors. Dispersive and 
magnetic effects will be neglected, whereas dissipation ($\eta$), inertia 
($\mu$) and elasticity ($\epsilon$) will be taken into account exactly in the 
general case of a finite number $N$ of layers. We shall call $d$ the 
inter-layer distance, $N$ the number of layers, $L$ the thickness of the 
film and $A$ its area. The continuum limit $N \rightarrow +\infty$ will be 
analyzed as a particular limit in which the usual model holds. 
The problem has already been addressed in the latter context by Blatter and 
Ivlev \cite{Blat4} for small magnetic fields and by Bulaevskii and Maley 
\cite{Bula1} for large magnetic fields. 

One of the central issues of this paper is to present a peculiar and
unambiguous relation between the value of the ratio $I = \frac {\sqrt{
\epsilon \mu}}{\eta d}$ and the 
thermodynamical properties of the vortex line. We show that there is a
temperature $T_o = \frac \pi 2 \frac \hbar {k_B} \frac \epsilon {\eta L^2}$,
typically of the order of a few $mK$, which separates the universal 
$I$-independent low temperature already discussed by Blatter and Ivlev 
\cite{Blat4} from an $I$-dependent higher temperature (but still
below the critical temperature) behavior.
In the relevant overdamped case in which $I \ll 1$, we find that in this
regime the specific heat is proportional to the square root of the
temperature. This should hold up to the critical temperature, provided
that the model is still appropriate. In Section IV we summarize our 
results; we compare them with some previous results appeared in the
litterature and discuss their physical relevance. In section V we discuss 
the effect of including the Magnus force. We find essentially the same 
results as before, in particular the same behavior with temperature, and even
the numerical values of the coefficient are not very different.

The fundamental characteristics of the vortex lines are quite simple in
the approximation we are considering; in particular, the self-energy
stemming from the interaction between neighbor pancake vortices essentially
reduces to a local elastic term, and dissipation seems to be reasonably well
described by an ohmic viscous term. Both the elastic modulus $\epsilon $ and
the friction coefficient $\eta $ can be theoretically computed, but the 
mass density $\mu $ is still object of discussion and can be only hardly 
estimated because of the wide class of contributions it gets depending on 
the particular setting, and its determination is still a central issue of 
vortex physics, even if there is strong belief for it to be negligibly small 
with respect to dissipation \cite{Stephen}. 

Since each point of the string has two-dimensional dynamics, we parameterize
the problem in cylindrical coordinates, calling ${\bf q}(t,z)$ the
xy position of the string at given z. The equation of motion of the vortex 
line is 
\begin{equation}
\label{eqnmot}
\mu {\bf \ddot q}+\eta {\bf \dot q}-\epsilon {\bf q^{\prime \prime }}=0\;. 
\end{equation}
Because of the finite size of the system in the $z$ direction, it is
necessary to impose the additional Neumann boundary conditions 
\begin{equation}
{\bf q}^{\prime }(t,0)={\bf q^{\prime }}(t,L)=0\;. 
\end{equation}
In this way, no energy-momentum flow is allowed across the end-points of the
string.

The quantum-mechanical version of this dissipative dynamics is reached by
means of Caldeira and Leggett's formalism \cite{Cald}, in which dissipation
is introduced by coupling the system to a thermal bath of harmonic
oscillators with a continuous frequency spectrum satisfying the requirement
to reproduce a simple velocity dependent ohmic viscous term in the effective
equation of motion. The Euclidean effective action of the open system,
relevant for thermodynamics, is obtained by integrating out the degrees of
freedom of the bath, and reads in the continuum limit 
\begin{eqnarray}
\label{effact}
S(\beta )=&&\int_0^Ldz \int_0^\beta d\tau \left[ \frac 12\mu 
{\bf \dot q}^2(\tau, z)+\frac 12\epsilon {\bf q^{\prime }}^2(\tau
, z)\right] \medskip\ \\ && +\frac \eta {4\pi
}\int_0^Ldz \int_0^\beta d\tau \int_0^\beta d\tau ^{\prime }\left[
\frac{{\bf q}(\tau , z)-{\bf q}(\tau ^{\prime }, z)}{
\frac \beta \pi \sin \left( \pi \frac{\tau -\tau ^{\prime }}
\beta \right)} \right]^2 \nonumber \;.  
\end{eqnarray}

It is convenient to extend ${\bf q}(t,z)$ symmetrically to negative $z$
, and expand the obtained even function in Fourier modes of fundamental
frequency $\nu =\frac \pi L$ along $z$ by means of ${\bf q}(t,z)={\bf q}
_0(t)+\sqrt{2}\sum_{n=0}^{N-1}{\bf q}_n(t)\cos \nu nz$; in this way, the
boundary conditions are automatically satisfied. The Euclidean action then
splits into a sum of harmonic oscillator contributions encoding the
vibrational modes of the string plus a zero mode representing the
translational motion of the center of mass.

In order to handle the thermodynamical Euclidean path-integral, we introduce
the Matsubara frequency $\omega =\frac{2\pi }\beta $ and expand the
periodic configurations ${\bf q}_n(\tau )$ involved in the path-integral for
the partition function in Fourier modes with fundamental frequency $\omega $
along imaginary time, taking ${\bf q}_n(\tau )=\frac 1{\sqrt{\beta }
}\sum_{k=-\infty }^{+\infty }{\bf q}_{nk}e^{i\omega k\tau }$. The
dissipative term, once Fourier-transformed, will contribute a peculiar $
\left| \omega k\right| $ term, remnant of the derivative nature of ohmic
dissipation.

In Fourier space, the effective action is 
\begin{equation}
\label{Action}S(\beta )=\frac 1 2 \sum \limits_{n=0}^{N-1}\sum
\limits_{k=-\infty }^{+\infty }\left[ M\left( \omega_k^2+\Omega_n^2\right) 
+ \Lambda \left| \omega_k\right| \right]
\left| {\bf q}_{nk}\right|^2 \;. 
\end{equation}
$M=\mu L$ is the total mass, $\Lambda = \eta L$ the total friction coefficient
and $\Omega = \sqrt{\frac \epsilon \mu} \frac \pi L$ the characteristic
vibration frequency;
we have used the notation $\omega_k = \omega k$ and $\Omega_n = \Omega n$.
Since there is only a finite number $N$ of layers, we have cut-off
the possible modes at that value.

A more natural way back to the finite $N$ case is to write a
discrete action in terms of the single pancake vortices coordinates ${\bf x}
_l(t)={\bf q}(t ,ld)$. One then obtains the action for $N$ pancake
vortices of mass $\mu d$ and friction coefficient $\eta d$, each
harmonically coupled to its nearest neighbors with elastic modulus $\frac
\epsilon d$. The decoupling of these degrees of freedom can be achieved
performing a change of variable analogous to the Fourier expansion of the
continuum case, defining ${\bf q}_n(t)$ from ${\bf x}_l(t)$ through $
{\bf x}_l(t)= {\bf q}_o(t) + \sqrt{2}\sum_{n=1}^{N-1}{\bf q}_n(t)
\cos \left( \nu n(l+\frac 12)d\right) $
. One then obtains the action (\ref{Action}), but
with a modified frequency spectrum given by 
\begin{equation}
\Omega _n=\frac{2N}\pi \sin \left( \frac \pi 2\frac nN\right) \Omega \;. 
\end{equation}
Obviously, in the limit of $N \rightarrow +\infty $, the lattice version of
the theory reduces to the continuum formulation, with the same frequency
spectrum $\Omega _n=\Omega n$.

In the following, for the sake of clearness we will concentrate on the basic
case of absence of pinning of vortex lines by impurities and defects, and 
focus on the effects of elasticity and dissipation (the effect of the 
Magnus force will instead be considered in Section V). However, the treatment 
of this important generalization could be easily faced in exactly the same 
way; introducing a single columnar pinning center through a harmonic 
confining potential with elastic modulus $k$ and characteristic frequency 
$\Omega_p=\sqrt{\frac k m}$, the only novel feature would be the modification 
of the spectrum of the modes to $\Omega_n^*=\sqrt{\Omega_n^2 + \Omega_p^2}$.

\section{Partition function}

We have seen that the dynamics of the vortex line is encoded in a zero
mode
describing its translational motion and $N-1$ harmonic modes with frequency
spectrum $\Omega _n$ describing its vibration. Consequently, the partition
function, 
\begin{eqnarray}
Z(\beta )&=&Tre^{-\beta H} \nonumber
\medskip\  \\ &=&\int_{{\bf q}(0,z)={\bf q}(\beta ,z)}{\cal D}{\bf q}
(\tau ,z)e^{-S\left( {\bf q}(\tau ,z)\right) }\;, 
\end{eqnarray}
will factorize into the product of the partition functions of the modes.
Thus, all that we need to know for our aim is the dissipative thermodynamics
of the zero-mode degree of freedom and the remaining harmonic
modes. The path-integral for the latter can be evaluated quite easily in
both cases since it is Gaussian, and reduces to a product of ordinary
integrals in Fourier space.

\subsection{Zero-mode}

For the zero mode, we get 
\begin{eqnarray}
Z(\beta ,\Lambda )&=& 
{\cal N}(\beta )\prod\limits_{k=0}^{+\infty }\int\!\!\!\int 
d{\bf q}_{ok}d{\bf q}
_{ok}^{*}e^{-{\bf q}_{ok}^{*}\left( M\omega_k^2+\Lambda \omega_k
\right) {\bf q}_{ok}}\nonumber \medskip \\ &=&{\cal N}^{\prime}(\beta
)A\prod\limits_{k=1}^{+\infty }\left( 1+\frac \gamma \omega \frac
1k\right) ^{-2}\; . 
\end{eqnarray}
where $\gamma =\frac \Lambda M=\frac \eta \mu $ is the characteristic
frequency related to dissipation and $A$ is the area. The normalization 
factor is fixed by
requiring that in the limit $\Lambda \rightarrow 0$ the partition function
reduces to the free one $Z_o(\beta )=\frac M{2\pi \beta }A$. It follows 
\begin{equation}
{\cal N}^{\prime}(\beta )=\frac M{2\pi \beta }\;. 
\end{equation}

Unfortunately, the infinite product representing the effect of dissipation
is divergent. We then have to introduce a frequency cut-off $\omega _c$
above which the effect of dissipation is assumed to be small, or at least
not adequately described by a simple ohmic viscous term in the equation of
motion, and truncate the product at $k_c(\omega )=\frac{\omega _c}\omega $.
Using the infinite product representation of the $\Gamma $-function,
Eq. (\ref{gam}), we obtain for $k_c(\omega )\rightarrow +\infty $, 
\begin{equation}
\prod\limits_{k=1}^{k_c(\omega )}\left( 1+\frac \gamma \omega \frac
1k\right) =\frac{k_c(\omega )^{\frac \gamma \omega }}{\frac \gamma \omega
\Gamma \left( \frac \gamma \omega \right) }\;. 
\end{equation}
The same result can be obtained using the Drude model as regularization \cite
{Weiss1,Weiss2}, where the frequency spectrum of the thermal bath is 
cut-off at $\omega _c$ dividing it by a factor $1+(\frac \omega {\omega _c})
^2$ ; the cut-off frequency can then be related to a microscopic relaxation 
time $\tau _c$ by $\omega _c=$ $\frac {2 \pi}{\tau _c}$.

The final form for the partition function is conveniently written in terms
of the function 
\begin{equation}
\label{gamtil}
\tilde \Gamma (z)=\frac 1{\sqrt{2\pi }}z^{\frac
12-z}e^z\Gamma (z) \;,
\end{equation}
which goes to one for large arguments away from the real
negative axis (see Eq. (\ref{andgamtil})); one gets 
\begin{equation}
Z(\beta ,\gamma )=\frac \Lambda {2\pi }A{\tilde \Gamma }^2\left( \frac
\gamma \omega \right) e^{-\beta {\cal E}_o(\gamma )}\;, 
\end{equation}
with a zero-point energy 
\begin{equation}
\label{eo}
{\cal E}_o(\gamma )=\frac \gamma \pi \left( 1+\ln \frac{\omega _c}\gamma
\right) \;. 
\end{equation}

Quite interestingly, we learn that, since the partition function depends
only on $\frac \omega \gamma $ apart from an irrelevant multiplicative
constant and the zero-point energy, the effect of dissipation on the
thermodynamics of the zero-mode is only to modify the temperature scale.
This could have been expected from dimensional analysis, since the only
energy scale we can construct from $M$ and $\Lambda $ is $\gamma $ (we do
not consider $\omega _c$ since we expect dissipation to manifest in a
universal ohmic way in the limit of small relaxation time $\tau _c$,
situation in which the only role of $\omega _c$ should be to renormalize the
energy of the open system), and the only dimensionless quantity that can
enter the partition function is the reduced temperature $\frac \omega
\gamma $.

Notice that in the limit of strong damping $\frac \gamma \omega \gg 1$, that
is equivalent to the small mass limit $\mu $$\rightarrow 0$ since $\frac
\omega \gamma =\frac{\mu \omega }\eta $, we are left, apart from the
zero-point energy, with a trivial constant partition function 
$Z_l(\beta, \gamma)$ independent of $\mu $; using Eq. (\ref{andgamtil})
\begin{eqnarray}
Z(\beta ,\gamma )
\longrightarrow\hspace{-20pt}\raisebox{-6pt}
{$\scriptscriptstyle{\frac \gamma \omega \gg 1}$}\;
&&\frac \Lambda {2\pi }A\left[ 1+\frac \omega {12\gamma }\right] ^2e^{-\beta 
{\cal E}_o(\gamma )} \nonumber \medskip\ \\ &&
\simeq \frac \Lambda {2\pi }Ae^{-\beta {\cal E}_o(\gamma
)}\;. 
\end{eqnarray}

\subsection{Harmonic modes}

For the other modes, we proceed exactly in the same way and the partition
function is given by 
\begin{eqnarray}
&&Z(\beta ,\Omega _n,\Lambda )= \nonumber
\medskip\  \\ &&\quad= 
{\cal N}(\beta)\prod\limits_{k=0}^{+\infty }\int\!\!\!\int d{\bf q}_{nk}d
{\bf q}_{nk}^{*}e^{-{\bf q}_{nk}^{*}\left[ M\left( \omega_k^2
+\Omega _n^2\right) +\Lambda \omega_k\right] {\bf q}_{nk}}\nonumber
\medskip \\ &&\quad=
{\cal N}^{\prime}(\beta)\frac 1{\Omega^2_n}
\prod\limits_{k=1}^{+\infty }\left[ 1+\frac
\gamma \omega \frac 1k+\left( \frac{\Omega _n}\omega \right) ^2\frac
1{k^2}\right] ^{-2}\;. 
\end{eqnarray}
The requirement this to reduce to the pure harmonic oscillator result 
$Z_o(\beta ,\Omega _n)=(2\sinh \frac{\beta \Omega _n}2)^{-2}$ in
the limit $\Lambda \rightarrow 0$ fixes, using Eq. (\ref{sinh}), the
normalization coefficient to 
\begin{equation}
{\cal N}^{\prime}(\beta)=\frac 1 {\beta^2}\;.
\end{equation}
As before, the dissipative term is responsible for a divergence in the
correction factor accounting for the damping of the system; introducing the
same frequency cut-off, the product can be carried out by
factorization, obtaining 
\begin{eqnarray}
&&\prod\limits_{k=1}^{k_c(\omega )}\left[ 1+\frac \gamma \omega \frac
1k+\left(\frac{\Omega _n}\omega \right) ^2\frac 1{k^2}\right] =
\nonumber \medskip \\ &&\qquad \qquad 
=\frac{k_c(\omega )^{\frac \gamma \omega }}{\left( \frac{\Omega _n}
\omega \right) ^2\Gamma \left( \frac{\frac \gamma 2+i\xi _n}\omega \right)
\Gamma \left( \frac{\frac \gamma 2-i\xi _n}\omega \right) }\;, 
\end{eqnarray}
where $\xi _n=\sqrt{\Omega _n^2-\frac{\gamma ^2}4}$. Again, the same result
can be obtained using the Drude model as regularization \cite{Weiss1,Weiss2},
allowing a microscopic interpretation of the cut-off.

The partition function can then be written, using again the 
$\tilde \Gamma$-function defined above, as 
\begin{equation}
Z(\beta ,\Omega _n,\gamma )={\tilde \Gamma }^2\left( \frac{\frac \gamma
2+i\xi _n}\omega \right) {\tilde \Gamma }^2\left( \frac{\frac \gamma 2-i\xi
_n}\omega \right) e^{-\beta {\cal E}_o(\Omega _n,\gamma )}\;, 
\end{equation}
with a zero-point energy given by 
\begin{equation}
{\cal E}_o(\Omega _n,\gamma )=\frac \gamma \pi \left( 1+\ln \frac{\omega _c 
}{\Omega _n}\right) -i\frac{\xi _n}\pi \ln \frac{\frac \gamma 2+i\xi _n}
{\frac \gamma 2-i\xi _n}\;. 
\end{equation}

In the strongly dissipated regime, which means $\frac \gamma \Omega \gg n$,
and successively in the limit of low temperature with respect to damping,
corresponding to $\frac \gamma \omega \gg 1$, the partition function
is seen, using respectively Eqs. (\ref{shigam}) and (\ref{andgamtil}),
to reduce to 
\begin{eqnarray}
Z(\beta ,\Omega _n,\gamma )
\longrightarrow\hspace{-20pt}\raisebox{-6pt}
{$\scriptscriptstyle{\frac \gamma \Omega \gg n}$}\;
&&{\tilde \Gamma }%
^2\left( \frac \gamma \omega \right) {\tilde \Gamma }^2\left( \frac{\Omega
_n^2}{\gamma \omega }\right) e^{-\beta {\cal E}_o(\gamma )}\medskip\  
\nonumber \\
\longrightarrow\hspace{-20pt}\raisebox{-6pt}
{$\scriptscriptstyle{\frac \gamma \omega \gg 1}$}\;
&&\left[ 1+\frac
\omega {12\gamma }\right] ^2{\tilde \Gamma }^2\left( \frac{\Omega _n^2}{%
\gamma \omega }\right) e^{-\beta {\cal E}_o(\gamma )}\medskip\ \nonumber  \\ 
&& \simeq {\tilde \Gamma }^2\left( \frac{\Omega _n^2}{%
\gamma \omega }\right) e^{-\beta {\cal E}_o(\gamma )}\;. 
\end{eqnarray}
The last extreme limit can be seen as a small mass limit $\mu \rightarrow
0 $. Since the variable $\frac{\gamma \omega }{\Omega ^2}=\left( \frac
N\pi \right) ^2\frac{\eta d^2}\epsilon \omega $ is independent of $\mu $,
the resulting behavior $Z_l(\beta ,\Omega _n,\gamma )$ no longer depends on 
$\mu $ apart from the zero-point contribution ${\cal E}_o (\gamma)$,
which is found to be the same as the one for the zero-mode, Eq. (\ref{eo}). 
Again, this is obvious from dimensional analysis. 
The only dimensionless quantity independent of $\mu $
we can form from the three variables $\omega $, $\gamma $ and $\Omega $ is 
$\frac{\gamma \omega }{\Omega ^2}$, indeed entering the resulting behavior;
instead, the $\mu$-dependent variable $\frac \omega \gamma = \frac{\mu
\omega }\eta $ disappears in the limit. Having eliminated one of the scales,
we fall in a regime in which, as for the zero-mode, dissipation
manifests itself only in the temperature scale.

\subsection{Vortex-string}

In order to compute the partition function of the string, all that we have
to do is multiply those relative to the modes that we have just computed,
that is 
\begin{eqnarray}
&&Z_{vor}(\beta ,\Omega ,\gamma )=Z(\beta ,\gamma
)\prod\limits_{n=1}^{N-1}Z(\beta ,\Omega _n,\gamma ) \nonumber 
\medskip \\ &&\qquad =\frac \Lambda {2\pi }A 
{\tilde \Gamma }^2\left( \frac \gamma \omega \right)
\prod\limits_{n=1}^{N-1}{\tilde \Gamma }^2\left( \frac{\frac \gamma 2+i\xi
_n}\omega \right) {\tilde \Gamma }^2\left( \frac{\frac \gamma 2-i\xi _n}
\omega \right) \medskip\ \nonumber \\ &&\qquad \quad \; 
e^{-\beta E_o(\Omega ,\gamma)}\;. 
\end{eqnarray}
The zero-point energy is the sum of those of the modes 
\begin{equation}
E_o(\Omega ,\gamma )={\cal E}_o(\gamma )+\sum\limits_{n=1}^{N-1}{\cal E}
_o(\Omega _n,\gamma )\;. 
\end{equation}

The first observation we can do is that for fixed $\omega $, $\gamma $ and 
$\Omega $, the factor $Z(\beta ,\Omega _n,\gamma )$ in the product over the
modes in the partition function goes to one for $n\rightarrow +\infty $, and
the latter is thus a well defined function of these variable for any $N$,
since the product can be shown to converge. The meaning of this observation
is that the variables we have chosen represent the true physical scales of
the problem, and increasing $N$ just add higher frequency modes that are
more and more suppressed on the fixed scale we consider.

Conversely, the zero-point energy increases indefinitely with $N$. Using the
lattice regularization and the results derived in appendix B, the total
zero-point energy can be recast in the following form 
\begin{equation}
E_o(\Omega ,\gamma )=E_{odiv}(\Omega ,\gamma )+E_{ofin}(\Omega ,\gamma
)\;, 
\end{equation}
with 
\begin{eqnarray}
&&E_{odiv}(\Omega ,\gamma )=\left\{ \frac N\pi \left( 
\frac{4N}\pi -1\right) \Omega -\frac 18\ln N\frac{\gamma ^2}\Omega
\right. \medskip\ \nonumber \\ &&\qquad \qquad \qquad \; \;
+\left. \left[ \frac N\pi \ln \left( \frac \pi N \frac{\omega _c}{\Omega}
\right)+\frac{\ln N}{2\pi }\right] \gamma \right\} \;, \medskip\ \\ 
&&E_{ofin}(\Omega ,\gamma )= \nonumber \medskip\ \\ && \qquad
=-\left\{ \frac 1{12}\Omega -\frac 1\pi \left[
1-\ln \left( \pi \frac \gamma \Omega \right) \right] \gamma +\frac 18\ln G 
\frac{\gamma ^2}\Omega \right\} \nonumber \medskip\  \\ &&\qquad \quad
-\sum\limits_{n=1}^{N-1}\left[ \Omega _n-\frac \gamma \pi 
-\frac 18\frac{\gamma ^2}{\Omega _n}+\ i\frac{\xi _n}\pi \ln 
\frac{\frac \gamma 2+i\xi _n}{\frac \gamma 2-i\xi _n}\right] \;. 
\end{eqnarray}
The last sum is finite also for $N \rightarrow + \infty$. Notice also that it 
vanishes in the free case $\gamma \rightarrow 0$.

In the limit in which each of the $N$ modes is in the strong damping regime
and at low temperatures with respect to friction, meaning $\frac \gamma 
\Omega \gg N$ and $\frac \gamma \omega \gg 1$, and corresponding to a limit
of vanishing mass $\mu \rightarrow 0$, the partition function first
factorizes essentially into the product of two functions depending
respectively on the combinations $\frac \omega \gamma =\frac{\mu
\omega }\eta $ and $\frac{\gamma \omega }{\Omega ^2}=\left( \frac N\pi
\right) ^2\frac{\eta d^2}\epsilon \omega $, and then simplifies to a
function $Z_{lvor}(\beta ,\Omega ,\gamma )$ which, apart from the
zero-point
energy, depends only on the $\mu $-independent variable $\frac{\gamma
\omega }{\Omega ^2}$. In fact, 
\begin{eqnarray}
\label{Zlim}
&&Z_{vor}(\beta ,\Omega ,\gamma ) \rightarrow \nonumber \medskip\ \\ && \qquad 
\longrightarrow\hspace{-20pt}\raisebox{-6pt}
{$\scriptscriptstyle{\frac \gamma \Omega \gg N}$}\;
\frac \Lambda
{2\pi }A{\tilde \Gamma }^{2N}\left( \frac \gamma \omega \right)
\prod\limits_{n=1}^{N-1}{\tilde \Gamma }^2\left( \frac{\Omega _n^2}{\gamma
\omega }\right) e^{-\beta E_o(\gamma )}\nonumber \medskip\  \\ &&
\qquad 
\longrightarrow\hspace{-20pt}\raisebox{-6pt}
{$\scriptscriptstyle{\frac \gamma \omega \gg 1}$}\;
\frac \Lambda
{2\pi }A\left[ 1+\frac \omega {12\gamma }\right]
^{2N}\prod\limits_{n=1}^{N-1}{\tilde \Gamma }^2\left( \frac{\Omega _n^2}{%
\gamma \omega }\right) e^{-\beta E_o(\gamma )} \nonumber \medskip\  \\ && 
\qquad \qquad \; \simeq \frac \Lambda {2\pi }A\prod\limits_{n=1}^{N-1}
{\tilde \Gamma }^2\left( \frac{\Omega _n^2}{\gamma \omega }\right) 
e^{-\beta E_o(\gamma)}\;, 
\end{eqnarray}
where now the zero-point energy has simplified to 
\begin{equation}
E_o(\gamma )=N{\cal E}_o(\gamma )=N\frac \gamma \pi \left( 1+\ln 
\frac{\omega _c}\gamma \right) \;. 
\end{equation}

Again, we see that the product in the partition function is well defined for
any $N$ since the $n$-th factor goes quickly to one for $n\rightarrow +
\infty$; the zero-point energy grows indefinitely if one add modes 
increasing $N$, as in the general case.

In the following, we shall distinguish between two regimes in the
thermodynamics of the vortex line: the underdamped one, for which
$\frac \gamma \Omega \ll N$, and the overdamped one, for which 
$\frac \gamma \Omega \gg N$.

\section{Specific heat}

Having computed the partition function of the vortex line, we can easily
compute its specific heat. The latter can be defined in presence of
dissipation as the derivative of the mean energy of the open system with
respect to temperature, considering dissipation just as an additional
mechanism for exchanging energy with the thermal bath and modifying the
response of the system under heating.

In fact, within the Caldeira-Leggett formulation, the partition function
of the complete system factorizes into the partition function of the
bath and that of the open system with an effective action containing
the influence of dissipation. Technically, this is implemented with an
appropriate shift in the bath's oscillators coordinates which are just
dummy variables in the path-integral \cite{Cald,Weiss1}. Thus, in this 
scheme, the total heat capacity of the system is the sum of the heat
capacity of the bath alone plus the one of the vortex lines. 
  
One of the interesting features of the specific heat is that it is
independent of the rather unknown parameter $\omega_c$ entering the
zero-point energy. Also, from our previous discussion, we expect it to
remain well behaved in the $\mu \rightarrow 0$ limit.

\subsection{General expression}

The specific heat of a single vortex line is 
\begin{equation}
C_{vor}(\beta , \Omega, \gamma )=-
\frac{\partial ^2}{\partial T\partial \beta }\ln Z_{str}
(\beta, \Omega, \gamma)\;.
\end{equation}
From now on, we will work with the reduced specific heat, dropping a 
factor $k_B$. The usual one is obtained by multiplying
the latter by $k_B$ and by the density of vortices in the sample.

The result for the specific heat is easily computed as a sum over the 
contributions of the modes and is found to be 
\begin{eqnarray}
\label{Cal}
&&C_{vor}(\beta ,\Omega ,\gamma )=C(\beta ,\gamma
)+\sum\limits_{n=1}^{N-1}C(\beta ,\Omega _n,\gamma ) 
\nonumber \medskip\  \\ &&\qquad \quad =2\left( \frac \gamma \omega \right)^2
\tilde \Psi ^{\prime }\left( \frac \gamma \omega \right) \nonumber
\medskip \\ &&\qquad \qquad +2\sum\limits_{n=1}^{N-1}\left\{ \left( 
\frac{\frac \gamma 2+i\xi _n}\omega \right) ^2\tilde \Psi ^{\prime }\left( 
\frac{\frac \gamma 2+i\xi _n}\omega \right) \right. \nonumber \medskip\  \\ 
&&\qquad \qquad \qquad \qquad 
+\left. \left( \frac{\frac \gamma 2-i\xi _n}\omega
\right) ^2\tilde \Psi ^{\prime }\left( \frac{\frac \gamma 2-i\xi _n}\omega
\right) \right\} \;, 
\end{eqnarray}
where
\begin{equation}
\label{psitilp}
\tilde \Psi^{\prime}(z)=\frac{d^2}{dz^2}\ln \tilde \Gamma (z)\;. 
\end{equation}
and goes to zero for large arguments away from the real negative axis,
Eq. (\ref{andpsitil}).
At high temperature, the specific heat saturates at the 
value $2N-1$; all the $N-1$ harmonic modes contributes $2$, 
whereas the zero-mode only $1$, as seen from Eq. (\ref{and2psitil}).

In the limit of strong damping for each of the $N$ modes, $\frac \gamma 
\Omega \gg N$, we have $\frac {\frac \gamma 2 \pm i \xi_n}\omega 
\rightarrow \frac {\Omega_n^2}{\gamma \omega}, \frac \gamma \omega$.
At low temperatures with respect to friction, $\frac 
\gamma \omega \gg 1$, the specific heat simplifies to 
\begin{eqnarray}
\label{Callim}
&&C_{vor}(\beta ,\Omega ,\gamma ) \rightarrow \nonumber \medskip\ \\ &&
\qquad \longrightarrow\hspace{-20pt}\raisebox{-6pt}
{$\scriptscriptstyle{\frac \gamma \Omega \gg N}$}\;
2N \left(\frac \gamma \omega \right)^2 \tilde \Psi^{\prime} \left(
\frac \gamma \omega \right) + 2\sum\limits_{n=1}^{N-1}\left( 
\frac{\Omega _n^2}{\gamma \omega }\right) ^2\tilde \Psi ^{\prime }\left( 
\frac{\Omega _n^2}{\gamma \omega }\right) \nonumber \medskip\  \\ &&
\qquad \longrightarrow\hspace{-20pt}\raisebox{-6pt}
{$\scriptscriptstyle{\frac \gamma \omega \gg 1}$}\;
\frac N3 \frac \omega \gamma + 2\sum\limits_{n=1}^{N-1}\left( 
\frac{\Omega _n^2}{\gamma \omega }\right) ^2\tilde \Psi ^{\prime }\left( 
\frac{\Omega _n^2}{\gamma \omega }\right) \nonumber \medskip\  \\ 
&& \qquad \qquad \; \simeq 2\sum\limits_{n=1}^{N-1}\left( \frac{\Omega _n^2}
{\gamma \omega } \right) ^2\tilde \Psi ^{\prime }\left( \frac{\Omega _n^2}
{\gamma \omega } \right) \;. 
\end{eqnarray}
In the final extreme limit, the specific heat reduces to a
function $C_{lvor}(\beta ,\Omega ,\gamma )$ which depends only 
on the $\mu $-independent variable $\frac{\gamma \omega }{\Omega ^2}$. 

\subsection{Behavior in various regimes}

As shown, the variables which naturally enter the thermodynamics of the
string have the dimensions of an energy and are the Matsubara frequency 
$\omega = 2 \pi k_B T$ which is proportional to the thermal energy, the 
characteristic frequency $\Omega = \sqrt{\frac \epsilon \mu} \frac \pi L$ 
of the string and $\gamma =\frac \eta \mu $ which is the characteristic 
energy of dissipative processes. 
These variables represent the dependence on temperature and damping on a
physical energy scale that depends on $L=Nd$ and thus on $N$, the 
number of layers.

It is convenient to define the reduced temperature $t$ and the 
dimensionless damping parameter $\alpha $ according to 
\begin{eqnarray}
&&t=\frac \omega \Omega = 2 N \sqrt{\frac \mu \epsilon} d k_B T =
\frac {\pi I}N \frac T{T_o} \;, \medskip\ \\
&&\alpha =\frac \gamma \Omega = \frac N \pi \frac {\eta d}
{\sqrt{\epsilon \mu}} = \frac N{\pi I}\;. 
\end{eqnarray}
The $\mu$-independent variable arizing in the $\mu \rightarrow 0$ limit 
is then simply 
\begin{equation}
\alpha t = \frac {\omega \gamma}{\Omega^2} = \frac {2 N^2}{\pi} 
\frac {\eta d^2}{\epsilon} k_B T = \frac T{T_o}\;.
\end{equation} 
 
The meaning of a possible continuum limit $N \rightarrow +\infty$  
can now be better elucidated. In order to compare situations with different 
$N$s, we work with the $N$-dependent variables $\alpha$ and $t$, so that 
adding modes does not change the contributions of those that were already 
present but constitute only a correction.
The continuum limit will then be relevant as a reliable approximation to the
finite $N$ case whenever the modes with $n>N$ give a negligible
contribution to thermodynamics. As we will see,
this will be true for low temperatures.

As we already pointed out, the thermodynamics of the vortex line clearly
exhibits two distinct and very different regimes.
We shall call the underdamped regime the case in which $\alpha \ll N$,
and the overdamped regime the case in which $\alpha \gg N$, and
analyze them separately.

\begin{figure}[h]
\centerline{\psfig{figure=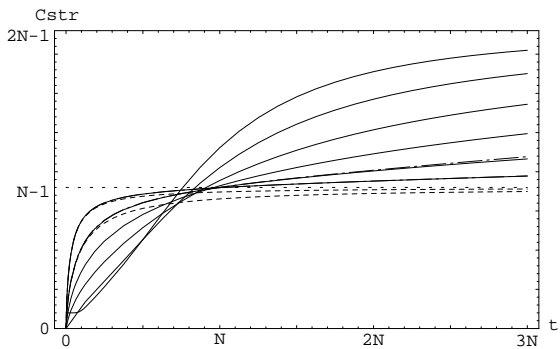,width=210pt}}
\caption{The specific heat $C_{vor}(t,\alpha )$ of the vortex-string as a
function of the reduced temperature $t$. The solid lines refers to 
$\alpha =$ 0, 3, $N$, $2N$, $4.5N$, $12N$. For $\alpha = 4.5N$, $12N$, the
limiting strong damping scaling behavior with (dot-dashed lines) and without 
(dashed lines) the linear correction (term $\frac N 3 \frac \omega \gamma$ in 
Eq.(\protect \ref{Callim})) is seen to be an increasingly good approximation.}
\label{fig1}
\end{figure}

The global behavior of the specific heat is shown in Fig. \ref{fig1}.  
There are two regimes according to the value of $\alpha $. 
For $\alpha \ll N$, the underdamped case, the specific heat depends 
only weakly on it and the effect of increasing $N$ is to raise the temperature 
at which the saturation to the asymptotic value $2N-1$ begins. 
For $\alpha \gg N$, the overdamped case, the specific heat acquires a 
strong dependence on $\alpha $. Increasing the temperature, the specific 
heat raises very quickly to the value $N-1$, going approximately like a 
square root, and continues to grow very slowly and almost linearly towards 
its high temperature saturation value $2N-1$.

In the limit of very small mass,
the specific heat stop its raising at $N-1$ instead of $2N-1$. This is due
to the fact that the extreme damping has killed almost all the kinetic
part of dynamics, corresponding to an asymptotic specific heat equal to $0$
instead of $1$ for the center of mass particle degree of freedom (which
gets completely killed) and equal to $1$ instead of $2$ for each of the
harmonic modes (for which the kinetic and potential part of the energy are
equal in the free case). Moreover, if damping is strong enough with respect 
to inertia, the specific heat depends only on the $\mu$-independent variable 
$\alpha t$.

\subsubsection{Underdamped regime: $\alpha \ll N$ ($I \gg 1$)}

In this regime, we use Eq. (\ref{Cal}), which depend on both $\alpha$ and 
$t$. Notice first that in the high temperature limit, $t \gg N$, the specific
heat tends to its maxmimal value $2N-1$ and is almost constant.
Next consider not too high temperatures, $t \ll N$. The contribution of 
the modes with $n>N$ in the sum giving the total specific heat would give a 
negligible contribution (since the arguments of both the $\tilde 
\Psi$-functions would be large), and the continuum limit is a good 
approximation. 
We can thus take the limit $N\rightarrow +\infty $ keeping $\alpha $ and $t$
fixed. In this way, the asymptotic behaviors of the specific heat in the 
extremes of this range of temperature can be computed and finally 
\begin{equation}
\label{Calund}
C_{vor}(t,\alpha )\simeq \left\{ 
\begin{array}{l}
\displaystyle{\frac 13\left( \frac 1\alpha +
\frac{\pi ^2}6\alpha \right) t\;,\;t\ll 1} \medskip\  \\ 
\displaystyle{\frac \pi 3t\;,\;1\ll t\ll N}
\medskip\  \\ 2N-1\;,\;t\gg N
\end{array}
\right. \;.
\end{equation}
The behavior for $t \ll 1$
directly follows from the asymptotic behavior of the $\tilde \Psi^{\prime}$-
function, Eq. (\ref{andpsitil}), whereas the one for $1 \ll t \ll N$ is
obtained approximating the sum in the specific heat with an integral.

At low temperatures, $t \ll N$, the underdamped case further splits
into three sub-regimes. For values of $\alpha $
below a first critical value $\alpha _{c1}\simeq 0.4$, the specific heat
progressively pass from the free shape to a linearly growing one; moreover,
any infinitesimal amount of dissipation, $\alpha >0$, causes the specific
heat to start from zero instead of one, because of the center of mass
contribution. For $\alpha $ beyond a second critical value $\alpha
_{c2}\simeq 1.5$, the specific heat starts to deviate from the linear shape,
getting more and more convex at low temperature. Finally, for $\alpha $
between the two critical values, there is no substantial dependence of the
specific heat on friction, and the response to heating is approximately
linear in the temperature in the whole continuum region $t \ll N$.

The two critical values of $\alpha$ can be obtained by requiring the matching 
of the behaviors for $t \ll 1$ and $1 \ll t \ll N$. They satisfy thus the
quadratic equation
\begin{equation}
\frac 13\left( \frac 1{\alpha _c}+\frac{\pi ^2}6\alpha _c\right) =\frac
\pi 3\;, 
\end{equation}
which yields
\begin{equation}
\alpha _{c1,2}=\frac{3\mp \sqrt{3}}\pi \;, 
\end{equation}
in agreement with the values extracted from the plot.

\subsubsection{Overdamped regime: $\alpha \gg N$ ($I \ll 1$)}

In this regime, we can use Eq. (\ref{Callim}), which depend only on
$\alpha t = \frac T{T_o}$.
At high temperature, $\alpha t \gg N^2$, the specific heat grows
linearly from the value $N-1$ with a slope $\frac N{3\alpha}$ until 
$\alpha t \sim \alpha^2$, where it starts saturating to its asymptotic
value $2N-1$. 
For low temperature limit, $\alpha t \ll N^2$, the contribution of 
the modes with $n>N$ in the sum giving the total specific heat are again 
negligible, and we can take the continuum limit $N\rightarrow +\infty$ 
keeping $\alpha t$ fixed. In this case, the asymptotic behaviors 
of the specific heat at the extremes of the low temperature region can be 
found and finally the behavior is the following
\begin{equation}
\label{Calover}
C_{vor}(t,\alpha )\simeq \left\{ 
\begin{array}{l}
\displaystyle{\frac{\pi ^2}{18}\alpha t\;,\;\alpha t \ll 1}
\medskip\  
\\ \displaystyle{2C_o \sqrt{\alpha t}\;,\;1 \ll \alpha t\ll N^2} 
\medskip\  \\ \displaystyle{
N-1+\frac N{3\alpha} t\;,\; N^2 \ll \alpha t \ll \alpha^2=\frac{N^2}
{\pi^2 I^2}} \medskip\ \\ 
\displaystyle{2N-1\;,\;\alpha t\gg \alpha^2 = \frac{N^2}{\pi^2 I^2}} 
\end{array}
\right. \;,
\end{equation}
where $C_o=0.490$ is the constant given by the integral (\ref{int})
quoted in appendix A.
As before, the behavior for $\alpha t \ll 1$ directly stems from the 
asymptotic behavior of the $\tilde \Psi^{\prime}$-function,
Eq. (\ref{andpsitil}), whereas the one for $1 \ll \alpha t \ll N^2$ is
computed approximating the sum over the modes with an integral.  

\vskip 25pt

The reduced temperatures that are important for the shape transitions in the 
specific heat are recognized to be $t=1,N$ in the underdamped case and
$t=\frac 1\alpha,\frac{N^2}\alpha, \alpha $ in the overdamped one.
The important question that we shall now address is whether some of these 
can be relevant at the superconductivity temperature scale.  

\section{Discussion}

The available experimental and theoretical work on superconductors and
vortex dynamics does not allow for a precise knowledge of all the parameters
entering the description of the problem. In particular, whereas the damping
and elasticity coefficients can be computed, the mass density is ambiguous.
We will thus focus on $I = \frac {\sqrt{\epsilon \mu}}{\eta d}$ as the
fundamental unknown quantity carrying the dependence on $\mu$. 

The value of $\eta$ can be estimated at low temperaure using the 
Bardeen-Stephen expression \cite{Bard}, whereas $\epsilon$ is known from the 
microscopic theory; in Gaussian units (and recovering $\hbar$), 
\begin{equation}
\eta = \frac{\Phi _o^2}{2\pi c^2\xi ^2\rho _N} \;,\;
\epsilon =\kappa^2 \left( \frac{\Phi _o}{4\pi \lambda }\right) ^2\ln 
\frac \lambda {\kappa \xi} \;, 
\end{equation}
where $\lambda $ is London's penetration depth, $\xi $ the $xy$ coherence
length, $\kappa = \sqrt{\frac mM}$ the anisotropy ratio, 
$\rho _N$ the normal state resistivity, and $\Phi _o=\frac{hc}{2e}$ the 
flux quantum.

For YBCO films, we can take the typical values $d \simeq 12\stackrel
{\circ }{A}$, $\lambda \simeq 1400\stackrel{\circ }{A}$, $\xi \simeq 
15\stackrel{\circ }{A}$, $\kappa \simeq \frac 15$ and 
$\rho_N \simeq 100\;\mu \Omega \cdot cm$, yielding 
\begin{equation}
\label{etaeps}
\eta \simeq 3.0\;10^{-6}\;\frac{Erg\cdot s}{cm^3} \;,\; 
\epsilon \simeq  3.4\;10^{-7}\;\frac{Erg}{cm}\;. 
\end{equation}

In order to get contact with realizable temperatures in the framework of
superconductivity, we define the $\mu$-independent temperature 
\begin{equation}
T_s=N T_o = \frac \pi {2 N} \frac \hbar {k_B} \frac \epsilon {\eta d^2}
\simeq \frac {10^2}{N}\;K \;. 
\end{equation}
For reasonable $N$, ranging from $10^2$ to $10^4$ and corresponding to a 
thickness $L$ of the sample between $0.1\;\mu m$ and $10\; \mu m$, $T_s$ 
can go from $10\;mK$ to $1\;K$, indeed representing the accessible 
temperature scale for the problem.
In the following, we will assume $T \ll \frac 1 {2\pi} \frac \hbar{k_B}
\frac \eta \mu$ in order to satisfy the condition $\frac \gamma \omega
\gg 1$ which has been used in deriving the behavior in the overdamped case.
Using the estimate given at the end of this section for 
$\mu$, this means temperatures below $10^5\;K$, and does not constitute any 
restriction.
    
In terms of $T_s$, the reduced temperature is given by 
\begin{equation}
t= \pi \frac {\sqrt{\epsilon \mu}}{\eta d} 
\frac T {T_s} \;. 
\end{equation}
Observe that since $T_s$ is known but $\mu$ is not, the temperature
scale entering $t$ is ambiguous.
We are now able to look closer to the order of magnitude of the transition
temperatures in the two regimes we have studied, taking as reasonable 
temperature scale $T_s$ ($T_o = \frac {T_s}N$ is thus small).

In the underdamped case, $\alpha= \frac N\pi \frac {\eta d}
{\sqrt{\epsilon \mu}} \ll N$, the important reduced 
temperatures where found to be $t_A=1$ and $t_B=N$, corresponding to 
the temperatures $T_A=\frac \alpha N T_s \ll T_s$ and $T_B = \alpha T_s 
\sim T_s$. Thus, we conclude that the relevant regimes for the specific 
heat are in this case the first and second of Eq. (\ref{Calund}), that is 
\begin{equation}
\label{Res1}
C_{vor} \simeq \left\{
\begin{array}{l}
\displaystyle{\frac N3 \left(\frac 1{\alpha^2} + \frac {\pi^2}6 \right)
\frac T{T_s}\;,\;T \ll \frac {T_s}N} \medskip\  
\\ \displaystyle{\frac \pi 3 \frac N \alpha \frac T{T_s}
\;,\;T \gg \frac {T_s}N}
\end{array}
\right. \;.
\end{equation} 

In the overdamped case, $\alpha=\frac N\pi \frac{\eta d}{\sqrt{\epsilon
\mu}} \gg N$, we had instead $t_A = \frac 1\alpha$,
$t_B = \frac {N^2}\alpha$ and $t_C = \alpha$, corresponding to the 
temperatures $T_A = \frac 1N T_s \ll T_s$, $T_B = N T_s \gg T_s$ and
$T_C = \frac {\alpha^2}N T_s \gg T_s$. Thus, the relevant regimes for this
case are the first and second of Eq. (\ref{Calover}), that is
\begin{equation}
\label{Res2}
C_{vor} \simeq \left\{
\begin{array}{l}
\displaystyle{\frac {N \pi^2}{18}
\frac T{T_s}\;,\;T \ll \frac {T_s}N} \medskip\  
\\ \displaystyle{0.98 \sqrt{N \frac T{T_s}}
\;,\;T \gg \frac {T_s}N}
\end{array}
\right. \;.
\end{equation} 

Multiplying the results (\ref{Res1}) and (\ref{Res2}) by $k_B$ and the 
density of vortices $\frac {B}{\Phi_o L}$ in order to obtain the true 
specific heat per unit volume, and expliciting all the variables, our
final result can be written as 
\begin{equation}
\label{Res}
C_{vor} \simeq \left\{
\begin{array}{l}
\left\{\begin{array}{l}
\displaystyle{\frac \pi 9 \frac B {\Phi_o} 
\frac {k_B^2}{\hbar} \left(\frac {\eta L}{\epsilon} +
\frac {6 \mu}{\eta L} \right) T \;,\;T \ll T_o} \medskip\  
\\ \displaystyle{\frac {2 \pi}3 \frac B {\Phi_o}
\frac {k_B^2}{\hbar} \sqrt{\frac \mu \epsilon} T
\;,\;T \gg T_o}
\end{array} \right. ,\;I \gg 1 \medskip\ \\
\left\{\begin{array}{l}
\displaystyle{\frac \pi 9 \frac B {\Phi_o} 
\frac {k_B^2}{\hbar} \frac {\eta L}{\epsilon} T \;,\;T \ll T_o} \medskip\  
\\ \displaystyle{0.98 \sqrt{\frac 8 \pi} \frac B {\Phi_o} 
\frac {k_B^{\frac 32}}{\hbar^{\frac 12}} \sqrt{\frac \eta \epsilon}
\sqrt{T}\;,\;T \gg T_o}
\end{array} \right. ,\;I \ll 1
\end{array} \right. \!\!\!\!\!\!\!\!\!\!\!\!\!\!\!\!\!\!\!
\end{equation}
with
\begin{equation}
T_o = \frac \pi 2 \frac \hbar {k_B} \frac \epsilon {\eta L^2} \;,\;
I= \frac {\sqrt{\epsilon \mu}}{\eta d} \;.
\end{equation}

In a previous work by Blatter and Ivlev \cite{Blat4} the low temperature
regime for small magnetic fields was considered using the usual continuum 
description; this corresponds to the limit $N \rightarrow \infty$ in our 
description and temperatures $T \ll T_o$. In the anisotropic and elastic 
limit, Eq. (33) of that work for the specific heat reduces to 
\begin{equation}
C_{vor} \simeq \frac 23 \frac B {\Phi_o} \frac {k_B^2}\hbar \frac \eta
\epsilon 
\int_{k_{zmin}}^{\infty} \frac {dk_z}{k_z^2} \;.
\end{equation} 
Our finite $N$ case can be recovered with the identifications $k_z 
\rightarrow \nu n $, $dk_z \rightarrow \nu$ ($\nu = \frac \pi L$),
meaning $k_{zmin}=\frac \pi L$ and
\begin{equation}
\int_{k_{zmin}}^{\infty} \frac {dk_z}{k_z^2} \rightarrow \frac 1 \nu 
\sum_{n=1}^{N-1} \frac 1{n^2} \simeq \frac \pi 6 L \;.
\end{equation}
In this way, one matches the linear and $\mu$-independent results of the 
first and third row of (\ref{Res}), which are relevant for $T \ll T_o$ and 
coincide in the continuum limit on wich we are focusing.

Note that out result Eq. (\ref{Res2}), case $T \gg \frac {T_s}N$, would
be modified by the introduction of a lower cut-off $k_{zmin}$ in the
following way
\begin{equation}
C_{vor} \simeq 2 \sqrt{N \frac {T}{T_s}} \int_{\sqrt{N \frac T{T_s}}
\frac d \pi k_{zmin}}^{\sqrt{N \frac {T}{T_s}}} dz z^4 \tilde \Psi^\prime
(z^2) \;.
\end{equation} 
We took $k_{zmin}=\frac \pi L$, but the behavior $C_{vor} \sim \sqrt{N
\frac T{T_s}}$ would hold also for other possible $k_{zmin}$, provided
$k_{zmin} \ll \frac \pi d \sqrt{\frac {T_s}{N T}}$. 

The behavior for large magnetic fields has been studied by Bulaevskii and
Maley \cite{Bula1}. In this case, the specific heat is still linear in the
temperature, but has a different dependence on the magnetic field and the
microscopic parameters of the vortices, and the result can not be expressed
as a function of only $\mu$, $\eta$ and $\epsilon$; this signals that in 
this regime the vortices do not retain their line structure \cite{Blat4} 
and other kind of configuration become important.

Summarizing, for $T \ll T_o$, and in the continuum limit $N \rightarrow 
\infty$, we find the universal behavior already studied in the litterature
\cite{Blat4}.
For $T \gg T_o$ and in the whole range of superconductivity temperatures 
we find instead a different behavior depending on the value of the parameter
$I$. In the underdamped case $I \gg 1$ the specific heat is linear in the 
temperature and depends only on inertia and not on friction, whereas in
the overdamped case $I \ll 1$ it goes like the square root of the 
temperature and depends only on friction and not on inertia. 

It is important to observe that for $N$ ranging from $10^2$ to $10^4$, the
temperature $T_o$ is very small and varies from $1\;\mu K$ to $10\;mK$.
This suggest that an interesting experimentally observable regime should 
be $T > T_o$.

As an example of theoretical estimate, one can approximate the mass
density $\mu$ with its electronic and electromagnetic contributions. 
These are found found to be \cite{Blat2,Suhl} 
\begin{equation}
\mu_{el} = \frac 2 {\pi^3} m_e k_F \;,\;
\mu_{em} = \left(\frac {\Phi_o}{4 \pi c \xi}\right)^2 \;.
\end{equation}
For YBCO films, the Fermi momentum is $k_F=0.5 \stackrel{\circ}{A^
{-1}}$, and one obtains $\mu_{el} \simeq 2.9 \; 10^{-21} \; \frac 
{gr}{cm}$, $\mu_{em} \simeq 1.2 \; 10^{-22} \; \frac {gr}{cm}$.
This would mean $I \simeq 0.1$, which does not corresponds clearly to
any of the two damping regimes. Thus, if one uses this theoretical estimate,
the thermodynamics should lie somewhere inbetween the underdamped and 
overdamped regime that we have considered in more detail. However, trusting
other conventional arguments \cite{Stephen}, the overdamped case should
be the relevant one. An experimental measure of the specific heat for 
$T \gg T_o$ could thus provide specific informations on this important
issue. 

\section{Effect of the Magnus force}

In this section we will discuss the influence of a possible Magnus force
on the thermodynamics and show how our results for the specific heat 
generalize to the case in which both the Magnus effect and friction are 
important.

The Magnus force corresponds to a term 
\begin{equation}
\delta \,^*\!{\bf \dot q}
\end{equation}
in the equation of motion (\ref{eqnmot}) of the vortex line, where 
$\,^*\!{\bf q} = {\bf q} \times {\bf z}$ is the dual of the $xy$
position. This term can be accounted for in the thermodynamics by adding
in the Euclidean effective action (\ref{effact}) the term
\begin{equation}
\int_0^Ldz \int_0^\beta d\tau \frac i2 \delta {\bf q} \,^*\!{\bf \dot q} \;.
\end{equation}
Actually, the Magnus force is rather controvertial \cite{Thou}. In the low 
magnetic field limit, by estimating $\delta$ from the number density of
the superconducting fluid, one would get a value of the same order of $\eta$
as in eq. (\ref{etaeps}).

Accordingly, the partition function for each of the modes of the vortex line
engoes the following modification
\begin{eqnarray}
&&\prod_{k=1}^{+ \infty}\left[1 + \frac \gamma \omega \frac 1k 
+ \left(\frac {\Omega_n}\omega \right)^2 \frac 1{k^2} \right]^{-2}
\rightarrow \nonumber \\ && \rightarrow 
\prod_{k=1}^{+ \infty}\left\{\left[1 + \frac \eta {\mu \omega} \frac 1k 
+ \left(\frac {\Omega_n}\omega \right)^2 \frac 1{k^2} \right]^2 
+\left(\frac {\delta}{\mu \omega}\right)^2 \frac 1{k^2}\right\}^{-1}
\nonumber \\ && \quad \;\; = 
\prod_{k=1}^{+ \infty}\left|1 + \frac {\hat \gamma} \omega \frac 1k 
+ \left(\frac {\Omega_n}\omega \right)^2 \frac 1{k^2} \right|^{-2}\;.
\end{eqnarray}
We have introduced the complex damping frequency
\begin{equation}
\hat \gamma = \frac {\eta + i \delta}{\mu} \;.
\end{equation}
This leads to the following modifications in the expressions for the 
partition functions
\begin{eqnarray}
&&\tilde \Gamma^2 \left(\frac \gamma \omega \right) \rightarrow
\tilde \Gamma \left(\frac {\hat \gamma} \omega \right)
\tilde \Gamma \left(\frac {\hat \gamma^*} \omega \right) \nonumber \\
&&\tilde \Gamma^2 \left(\frac {\frac \gamma 2 \pm i \xi_n}
\omega \right) \rightarrow
\tilde \Gamma \left(\frac {\frac {\hat \gamma}2 \pm i \hat \xi_n} 
\omega \right) \tilde \Gamma \left(\frac {\frac {\hat \gamma^*}2 
\pm i \hat \xi^*_n} \omega \right)
\end{eqnarray}
where
\begin{equation}
\hat \xi_n = \sqrt{\Omega_n^2 - \frac {\hat \gamma^2}4} \;\;,\;\;
\hat \xi^*_n = \sqrt{\Omega_n^2 - \frac {\hat \gamma^{*2}}4} \;.
\end{equation}

All the analysis done for the dissipative case goes through without
any major difficulty. Underdamped and overdamped cases now means with 
respect to the modulus $\eta_e$ of the complex friction coefficient
$\hat \eta = \eta_e e^{i\phi}$ for which
\begin{equation}
\eta_e = \sqrt{\eta^2 + \delta^2} \;\;,\;\;
\phi = \arctan \frac \delta \eta \;.
\end{equation}
The result (\ref{Res}) generalizes to
\begin{equation}
\label{Resmod}
C_{vor} \simeq \left\{
\begin{array}{l}
\left\{\begin{array}{l}
\displaystyle{\frac \pi 9 \frac B {\Phi_o} 
\frac {k_B^2}{\hbar} \left(\frac {\eta L}{\epsilon} +
\frac {6 \mu \eta}{\eta^2_{e} L}
\right) T \;,\;T \ll \hat T_o} \medskip\  
\\ \displaystyle{\frac {2 \pi}3 \frac B {\Phi_o}
\frac {k_B^2}{\hbar} \sqrt{\frac \mu \epsilon} T
\;,\;T \gg \hat T_o}
\end{array} \right. \!\!,\;\hat I \gg 1 \medskip\ \\
\left\{\begin{array}{l}
\displaystyle{\frac \pi 9 \frac B {\Phi_o} 
\frac {k_B^2}{\hbar} \frac {\eta L}{\epsilon} T \;,\;T \ll \hat T_o} 
\medskip\  
\\ \displaystyle{2C_{\phi} \sqrt{\frac 8 \pi} \frac B {\Phi_o} 
\frac {k_B^{\frac 32}}{\hbar^{\frac 12}} 
\sqrt{\frac {\eta_{e}} \epsilon}
\sqrt{T}\;,\;T \gg \hat T_o}
\end{array} \right. ,\;\hat I \ll 1
\end{array} \right. \!\!\!\!
\end{equation}
with
\begin{equation}
\hat T_o = \frac \pi 2 \frac \hbar {k_B} 
\frac \epsilon {\eta_{e} L^2} \;,\;
\hat I= \frac {\sqrt{\epsilon \mu}}{\eta_{e} d}\;.
\end{equation}

The constant $C_{\phi}$ appearing in the overdamped case now depends on the 
phase $\phi$ and is again given by an integral
\begin{equation}
C_{\phi} = \Re \left\{e^{-2i\phi}\int_{0}^{+\infty} dz z^4 \tilde \Psi^{\prime}
(e^{-i\phi}z^2)\right\}\; .
\end{equation}
Since the relevant asymptotic behavior of the integrand for $z \rightarrow 0$
is independent of $\phi$,
we expect $C_{\phi}$ to depend only weakly on it. In fact, the to extreme
values corresponding to $\eta \neq 0$, $\delta = 0$ ($\phi = 0$) and
$\eta = 0$, $\delta \neq 0$ ($\phi = \frac \pi 2$) are given by the integrals
(\ref{int}) and (\ref{int2}) of appendix A,
\begin{equation}
C_o = 0.490 \;\;,\;\; C_{\frac \pi 2} = 0.346 \;.
\end{equation}
For arbitrary $\phi$, $C_{\phi}$ is of the same order of magnitude and
can be computed numerically. 

Observe finally that in the overdamped case, in the sense explained above 
(with respect to $\eta_e$), the specific heat depends on the square root
of the temperature and on the effective damping coefficient 
$\eta_e = \sqrt{\eta^2 + \delta^2}$. In the ultra clean limit 
($\eta_e = \delta$) this was already noticed bye Blatter and Ivlev 
\cite{Blat4}, who also found $C_v \sim \sqrt{T}$ in this case.

{\bf Aknowledgements}. This work has been partially supported by EEC
contract ERBFMRXCT 96-0045.

\appendix 

\section{Properties of $\Gamma $-functions}

In this appendix, we recall some formulae involving $\Gamma $-functions and
other \cite{Grad}. First of all, there are the infinite product
representations of the hyperbolic sine and the $\Gamma $-function, given
respectively by 
\begin{eqnarray}
\label{sinh}
&&\sinh z= z \prod\limits_{k=1}^{+\infty }\left[ 1+\left( 
\frac{z}{\pi k}\right) ^2\right] \; ,\medskip\ \\
\label{gam}  
&&\Gamma (z)=\lim
\limits_{N\rightarrow +\infty }\frac{N^z}z\prod\limits_{k=1}^N\left(
1+\frac zk\right) ^{-1}\;. 
\end{eqnarray}

We recall the two useful formulae, holding away from the real negative axis, 
\begin{eqnarray}
\label{andgam}
&&\Gamma (z)
\hspace{4pt}\longrightarrow\hspace{-24pt}\raisebox{-6pt}
{$\scriptscriptstyle{\left| z\right| \rightarrow +\infty }$}\;
\sqrt{2\pi }z^{z-\frac 12}e^{-z}(1+\frac 1{12z}+...)\; ,\medskip\ \\
\label{shigam} 
&&\Gamma (z+x) \hspace{4pt}\longrightarrow\hspace{-24pt}\raisebox{-6pt}
{$\scriptscriptstyle{\left| z\right| \rightarrow +\infty }$}\;
z^x\Gamma (z) \; .  
\end{eqnarray}

Notice the simplified asymptotic behaviors of the functions defined in
Eqs. (\ref{gamtil}) and (\ref{psitilp}) 
\begin{eqnarray}
\label{andgamtil}
&&\tilde \Gamma (z) 
\hspace{4pt}\longrightarrow\hspace{-24pt}\raisebox{-6pt}
{$\scriptscriptstyle{\left| z\right| \rightarrow +\infty }$}\; 
1+\frac 1{12z}+...\simeq 1\; ,\medskip\  \\
\label{andpsitil} 
&&\tilde \Psi ^{\prime }(z)
\hspace{4pt}\longrightarrow\hspace{-24pt}\raisebox{-6pt}
{$\scriptscriptstyle{\left| z\right| \rightarrow +\infty }$}\;
\frac 1{6z^3}+...\simeq 0 \; . 
\end{eqnarray}

All the behavior for small arguments can be deduced from the one of the 
$\Gamma$-function which has a simple pole in the origin. In particular
\begin{eqnarray}
\label{and2gamtil}
&&\tilde \Gamma (z) 
\hspace{4pt}\longrightarrow\hspace{-20pt}\raisebox{-6pt}
{$\scriptscriptstyle{\left| z\right| \rightarrow 0}$}\;
\frac 1{\sqrt{2 \pi z}}\; ,\medskip\  \\
\label{and2psitil} 
&&\tilde \Psi ^{\prime }(z)
\hspace{4pt}\longrightarrow\hspace{-20pt}\raisebox{-6pt}
{$\scriptscriptstyle{\left| z\right| \rightarrow 0}$}\;
\frac 1{2z^2} \; . 
\end{eqnarray}

The $\tilde \Gamma$ and $\tilde \Psi$ functions are related to hyperbolic 
functions at immaginary arguments
\begin{eqnarray} 
&&\tilde \Gamma (ix) \tilde \Gamma (-ix) = \frac {e^{\pi x}}{2 \sinh \pi x}
\;,\medskip\ \\
&&\tilde \Psi^\prime (ix) + \tilde \Psi^\prime (-ix) = 
-\left(\frac {\pi}{\sinh \pi x}\right)^2\;.
\end{eqnarray}

Finally, we will need the following integrals
\begin{eqnarray}
\label{int}
&&C_o = \int_{0}^{+\infty} dz z^4 \tilde \Psi^{\prime}(z^2) = 0.490 \; ,
\medskip\ \\
\label{int2}
&&C_{\frac \pi 2}= \frac 12 \int_{0}^{+\infty} dz \left(
\frac {\pi z^2}{\sinh \pi z^2}\right)^2 = 0.346 \;.
\end{eqnarray}

\section{Lattice regularization}

We present here some results about divergent quantities computed with the
lattice regularization. Using the lattice spectrum 
\begin{equation}
\Omega _n=\frac{2N}\pi \sin \left( \frac \pi 2\frac nN\right) \Omega 
\end{equation}
and the formulae 
\begin{eqnarray}
&&\sum\limits_{n=1}^{N-1}2\sin \left(\frac \pi 2\frac nN \right)
=\cot \frac \pi {4N}-1\simeq \frac{4N}\pi -1-\frac \pi {12N}\; ,\medskip\ 
\nonumber \\ &&\prod \limits_{n=1}^{N-1}2\sin \left(\frac \pi 2\frac nN 
\right)= \sqrt{N}\; ,\medskip \\ 
&&\sum\limits_{n=1}^{N-1}\frac{\frac \pi {2N}}
{\sin \left( \frac \pi 2\frac nN \right)}\simeq \ln GN\;,\; G=2.267\; , 
\nonumber
\end{eqnarray}
it easily follows that 
\begin{eqnarray}
&&\sum\limits_{n=1}^{N-1}\Omega _n\simeq \left[ \frac N\pi \left( 
\frac{4N}\pi -1\right) -\frac 1{12}\right] \Omega \; ,\medskip\ \nonumber \\
&&\prod\limits_{n=1}^{N-1}\Omega _n=\left( 
\frac N \pi \Omega \right) ^{N-1}\sqrt{N}\; ,\medskip \\ 
&&\sum\limits_{n=1}^{N-1}\frac 1{\Omega _n}\simeq \frac 1\Omega \ln GN 
\nonumber \;. 
\end{eqnarray}

\end{document}